\documentclass[twoside,journey]{IEEEtran}
\usepackage{makecell}
\usepackage{hyperref}
\usepackage{array}
\usepackage{graphicx,amssymb,amsmath}
\usepackage{multicol}
\usepackage[noadjust]{cite}
\usepackage{setspace}
\usepackage{subfigure}
\usepackage{graphicx}
\usepackage{float}
\usepackage {url}
\usepackage{stfloats}
\usepackage{amsthm,pifont}
\usepackage{flushend}
\usepackage{cases,subeqnarray}
\usepackage{bm,multirow,bigstrut}
\usepackage{amsmath, amsthm, amssymb}
\usepackage{textcomp}
\usepackage{latexsym,bm}
\usepackage{booktabs}
\usepackage{xcolor}
\usepackage{mathtools}
\usepackage{dsfont}
\usepackage{extarrows}
\usepackage{epsfig}
\usepackage{epsfig}
\usepackage{epstopdf}
\usepackage[noend]{algpseudocode}
\usepackage{algorithmicx,algorithm}
\usepackage{xspace}
\usepackage{listings}
\usepackage{xcolor}
\theoremstyle{plain}

\theoremstyle{plain}

\usepackage{amsmath}

\IEEEoverridecommandlockouts
\begin{document}
\title{Rethinking Quality of Experience for Metaverse Services: A Consumer-based Economics Perspective}
\author{Hongyang Du, Bohao Ma, Dusit Niyato,~\IEEEmembership{Fellow,~IEEE}, Jiawen Kang, Zehui Xiong, and Zhaohui Yang
	\thanks{H.~Du and D. Niyato are with the School of Computer Science and Engineering, Nanyang Technological University, Singapore (e-mail: hongyang001@e.ntu.edu.sg, dniyato@ntu.edu.sg).}
	\thanks{B.~Ma is with the Shool of Civil and Environmental Engineering, Nanyang Technological University, Singapore (e-mail: bohao001@e.ntu.edu.sg).}
	\thanks{J. Kang is with the School of Automation, Guangdong University of Technology, China (e-mail: kavinkang@gdut.edu.cn).}
	\thanks{Z. Xiong is with the Pillar of Information Systems Technology and Design, Singapore University of Technology and Design, Singapore (e-mail: zehui\_xiong@sutd.edu.sg).}
	\thanks{Z. Yang is with the College of Information Science and Electronic Engineering, Zhejiang University, China (e-mail: yang\_zhaohui@zju.edu.cn).}
}
\maketitle
\vspace{-1cm}
\begin{abstract}
The Metaverse is considered to be one prototype of the next-generation Internet, which contains people's expectations for the future world. However, the academic discussion of the Metaverse still mainly focused on the technical system design, and few research studied Metaverse challenges from the perspective of consumers, i.e., Metaverse users. One difficulty is that the analysis from the consumer's perspective requires interdisciplinary theoretical framework and quantifiable Quality of Experience (QoE) measurements. In this article, pioneering from consumers' point of view, we explore an interaction between Metaverse system design and consumer behaviors. Specifically, we rethink QoE and propose an framework that encompasses both the Metaverse service providers (MSPs) and consumer considerations. From the macro perspective, we introduce a joint optimization scheme that simultaneously considers the Metaverse system design, consumers' utility, and profitability of the MSPs. From the micro perspective, we advocate the Willingness-to-Pay (WTP) as an easy-to-implement measurement for future Metaverse system studies. To illustrate the usability of the proposed integrated framework, a use case, i.e., virtual traveling, is presented. We show that our framework can benefit the MSPs in offering competitive and economical service design to consumers while maximizing the profit.
\end{abstract}
\IEEEpeerreviewmaketitle
\section{Introduction}
\IEEEPARstart{S}{upported} by the development of communication technologies, the Metaverse is an emerging concept considered to be the next generation of the Internet. Instead of accessing the Internet through mobile phones and computers, users can completely immerse themselves in the virtual world of the Metaverse through interactive technologies such as Virtual Reality (VR) and Augmented Reality (AR). In addition, Metaverse can provide support for the construction of the real world. For example, complex experiments that are difficult to be performed in the real world can be simulated in the Metaverse, and then the results are fed back to the real world for decision-making. These desirable features have drawn the attention of giant multinational companies globally. For instance, {\textit{Facebook}}, which recently changed its name to {\textit{Meta}}, announced ambitious the plan to invest $\$10$ billion in Metaverse development. {\textit{Microsoft}} acquired company {\textit{Activision Blizzard, Inc.}} for $\$70$ billion to support its Metaverse strategy.

To exploit the full potential of the Metaverse, several system design indicators have been proposed and widely adopted, e.g., high-reliability feedback, high-definition virtual object perception, and feeling of immersion. Both the industry and academia have made numerous efforts to improve the performance based on the proposed indicators. For example, the development of computational graphic software such as Unreal Engine has made it possible to render scenes that are close to places of interest in the real world. In addition, advances in hardware, such as head-mounted devices (HMDs), can provide users with high-quality virtual scene perception over long periods of time to create an immersive experience in the Metaverse. Meanwhile, it is foreseeable that Extended Reality (XR) devices in the near future can further improve the quality of immersion. Massive data transmission can be efficiently supported by advances in communication technologies, e.g., massive multi-input multi-output (MIMO) technology, reconfigurable intelligent surfaces, mmWave, and semantic communication techniques. Beyond fifth-generation (B5G) and the sixth-generation (6G) wireless communication network can provide a huge number of Metaverse devices with real-time, ubiquitous, seamless, and ultra-reliable communications~\cite{tataria20216g}.

\begin{figure*}[t]
	\centering
	\includegraphics[width = 0.83\textwidth]{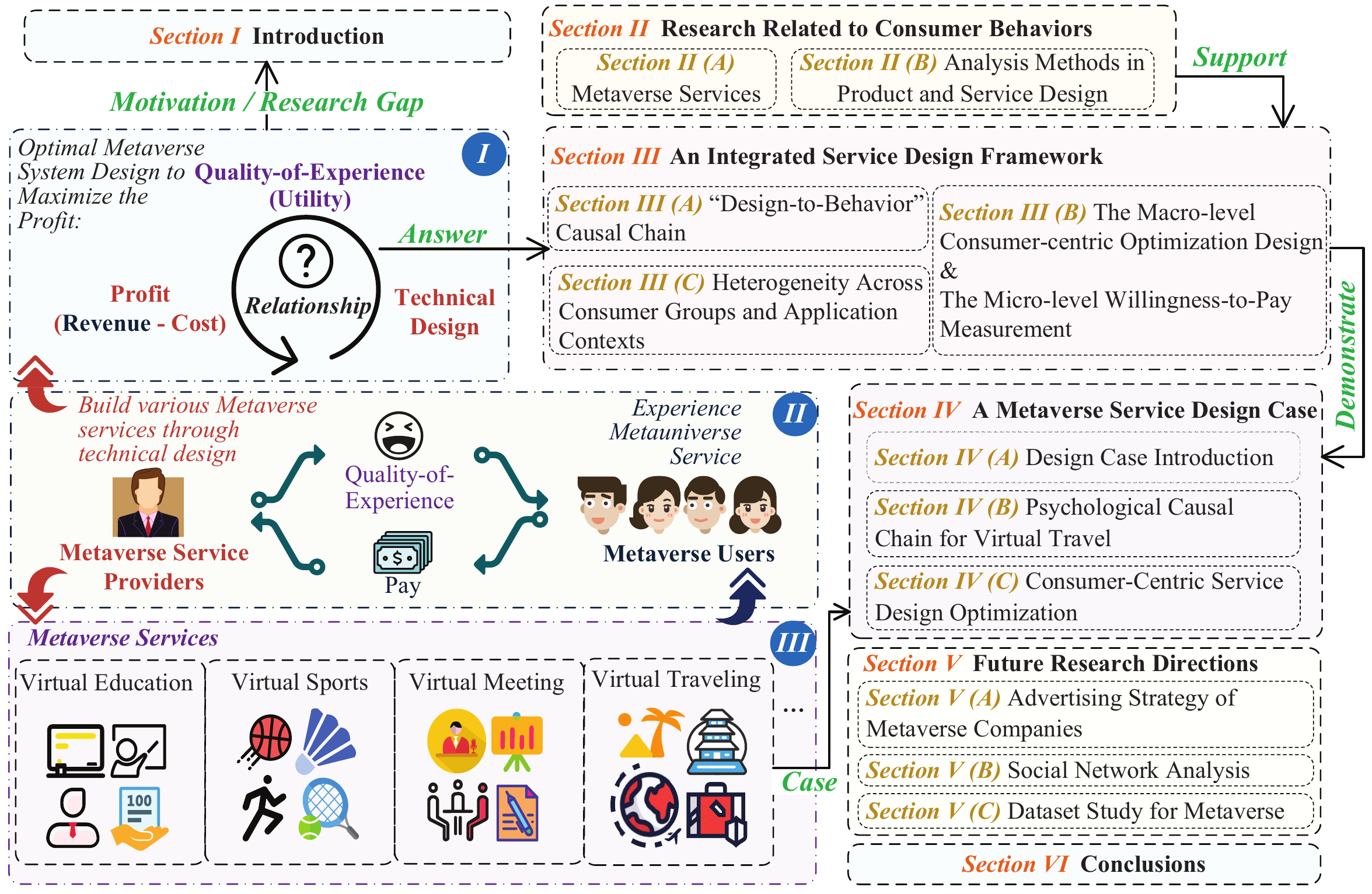}
	\caption{Structure of this article.}
	\label{Visio-Structure} 
\end{figure*}

Besides the technical development, a conceptual shift required by the Metaverse is the adoption of a user-centric perspective throughout the entire design process. This shift can also be observed in the new generation Internet, e.g., Web3. Unlike ``read based'' Web1 and ``read and write based'' Web2, Web3 is based on ``read, write, and own''. Although the success of the Metaverse critically depends on active user participation, in the current development of the Metaverse and related technologies, consumers as the vital success factor, have received little attention. Attempts by Metaverse service providers (MSPs) to enhance user-centric design are facing many difficulties. The reason is that, unlike the objective and directly measurable technical indicators, the interaction between system design with consumer behaviors is rather complex, as shown in Fig.~\ref{Visio-Structure} (Part I). Thus, it is unclear how the consumer perspective can be incorporated into the Metaverse system design process. Several studies attempted to propose Quality-of-Experience (QoE) measurements for the user-centric design in the Metaverse~\cite{wang2016data}. Yet the proposed indicators are considered to be highly simplified and may not capture the complex interaction between consumer behavior and system designs. Considering that the Business-to-Customer (B2C) model will play an important role in the promotion of Metaverse, behavior analysis from consumer perspectives is essential for the MSPs.

With all above motivation, in this article, we propose an interdisciplinary framework that qualitatively synergies the technical and consumer perspectives from both macro-level (system-level), and micro-level (individual-level). Specifically, the contributions of this article can be summarized as follows.
\begin{itemize}
    \item We explore the causal relationship between Metaverse system design and consumer behaviors with the support of consumer economic research. A Stimulus-Organism-Response (S-O-R) behavioral framework is established.
    \item We propose the implementation of econometric discrete choice model (DCM) as the macro-level and a Willingness-to-Pay (WTP) measure as the micro-level to connect the system design with consumer behaviors.
    \item Taking a specific Metaverse service, i.e., virtual traveling, as a case study, we demonstrate how to use our proposed framework to jointly consider the Metaverse system design, consumers' utility and profitability of the MSPs.
\end{itemize}
The structure of the article is shown in Fig.~\ref{Visio-Structure}. We discuss the existing system design research focusing on the consumer-centric concept, and then propose our integrated Metaverse design framework. Using a case study, we demonstrate the applicability of the framework. Finally, we discuss possible future research directions and conclude this article.

\section{Research Related to Consumer Behaviors}\label{S2}
In this section, we first introduce the analysis methods of consumer behaviors in existing Metaverse research with their limitations. Then, we present how user behavior is generally considered in the design of products and services.

\subsection{Metaverse Services}
The progress of multimedia technologies, e.g., VR and AR, and digital economy technologies, e.g., blockchain, has promoted the emergence of Metaverse services. Although many Metaverse services are still in their infancy, we can observe that the services have shown a diversified evolution direction. For example, MSPs can provide virtual educational, traveling, game, and teleconferencing services, as shown in Fig.~\ref{Visio-Structure} (Part III). To provide formal guidance for the Metaverse system, quality of service (QoS) and QoE are widely used as utility functions. Specifically, the QoS attempts to measure objectively system performance metrics, i.e., data transmission rate and bit error probability. QoE is a subjective measure from the perspective of the user~\cite{wang2016data}.

However, it is not realistic to consider QoS or QoE as an objective function in the design of Metaverse systems directly and blindly. Although these two indicators are positively correlated to the profit of MSPs in most cases, the relationship between the profit and the utility, i.e., QoS or QoE, is actually complicated. When a MSP invests more resources in the system design to improve QoE and QoS, the MSP needs to raise prices to ensure its profitability~\cite{du2022attention}. However, different users have different sensitivities to the rising service prices, i.e., users are heterogenous instead of homogeneous. Moreover, although the increase in QoE can bring the increase in the loyalty of existing users and the increase in the number of new users, the specific relationships are difficult to quantify. Thus, most of the current literature ignores complex user behavior responses by various assumptions and simplifications. To fill this research gap, we aim to rethink QoE measurement and design a comprehensive utility function analysis framework for MSP by considering consumer behavior in depth in the Metaverse system design.

\subsection{Analysis Methods in Product and Service Design}
 \begin{figure}[t]
    \centering
    \includegraphics[width = 0.42\textwidth]{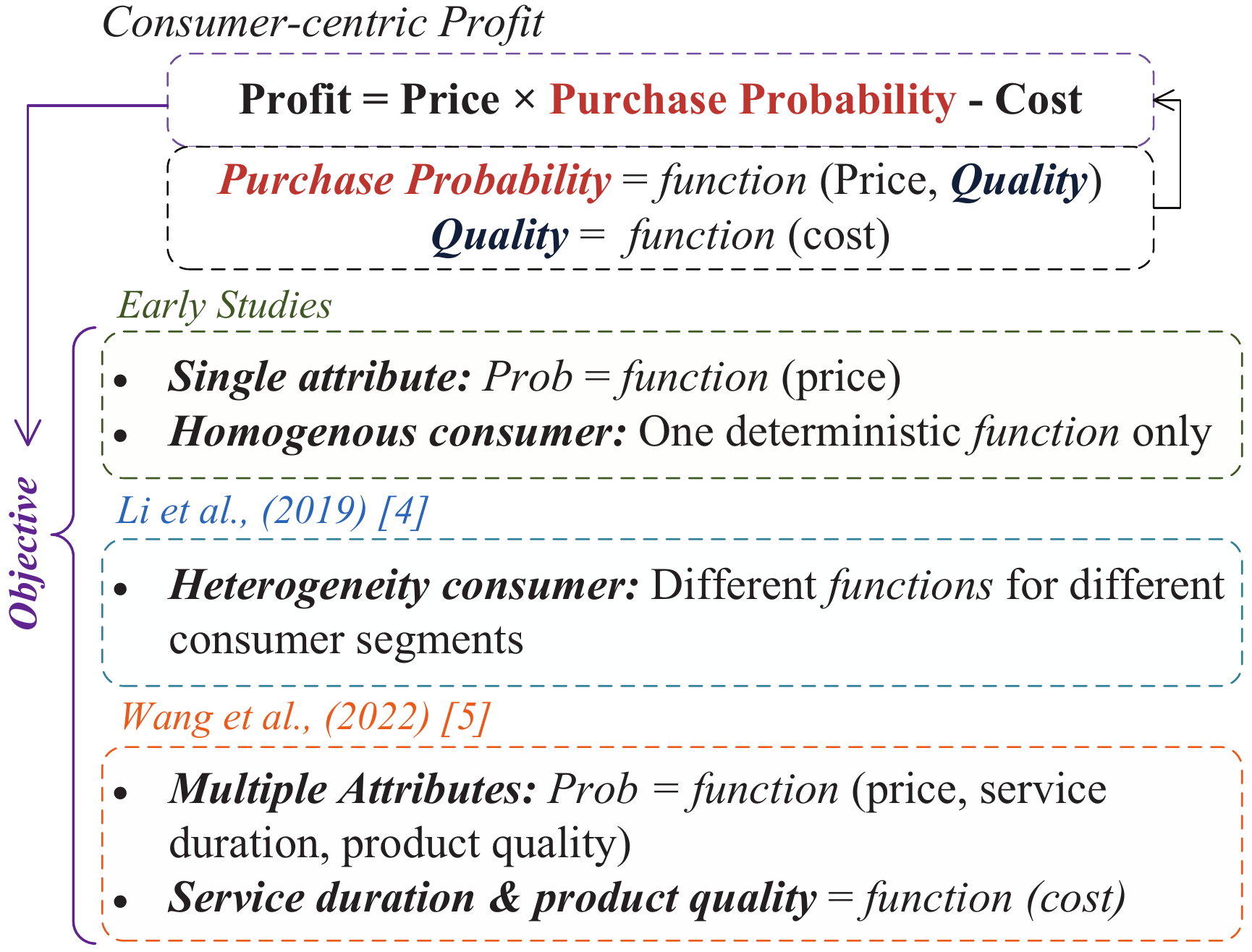}
    \caption{Illustration of consumer behaviors in product/service design.}
    \label{Visio-fig3} 
    \end{figure}
In the literature on product line design and hospitality management, the importance and value of incorporating consumer behaviors have long been noted. 
As shown in Fig.~\ref{Visio-fig3}, in the early years, research was limited to single-attribute design, namely pricing optimization. In addition, consumers are assumed to be homogeneous. Yet, significant progress has been made in recent years, as a result of both advancements in mathematical modeling and improvement in computational powers. In a product line design case, the authors in \cite{li2019product} formulate a pricing problem with consumers' heterogeneity, and apply a latent-class multinomial formulation. The authors assume that consumers hold different views towards product attributes. Such heterogeneity can be described with tailored models for each consumer. For example, because consumers' sensitivity to prices may vary, the whole population can be segmented into high-income, medium-income and low-income groups. Within each group, specific price sensitivity is applied. The latent-class-based model was applied to a real-life consumer electronic pricing optimization case, and significant performance improvement of the manufacturer was observed. 
From another perspective, the authors in \cite{wang2022product} studied an integrated product design problem whereby a firm jointly determines multiple product attributes, e.g., price, quality and service duration. It is shown that such product differentiation can both benefit the consumers and improve the profitability of the firm.

One observation is that the most notable success in the above analysis is made by the multinomial formulation based on the Random Utility Model \cite{train2009discrete}. Such a formulation supports design decisions by quantifying the impacts of product and service attributes on consumers' behaviors. More details on applying the Random Utility Model in Metaverse system design are given in Section~\ref{S32}. 


\section{An Integrated System Design Framework}\label{S3}
\begin{figure}[t]
	\centering
	\includegraphics[width = 0.48\textwidth]{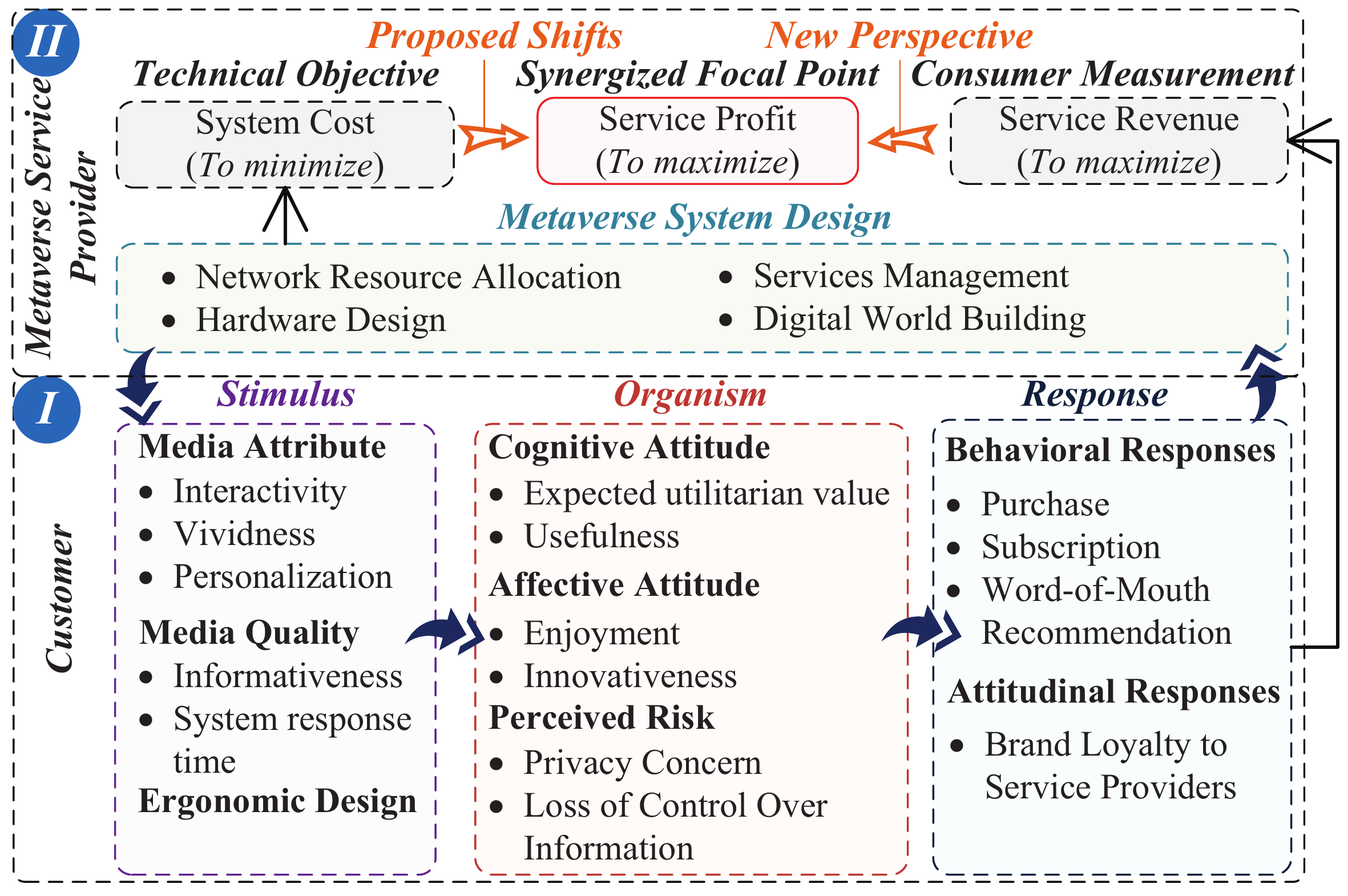}
	\caption{Stimulus-Organism-Response (S-O-R) behavioral framework.}
	\label{Visio-fig1} 
\end{figure}
In this section, we propose an integrated framework that synergizes the response of consumers and Metaverse system design. We first introduce the design-to-behavior causal chain, and then connect the system design with consumer behaviors.

\subsection{The Design-to-Behavior Causal Chain}\label{S31}
In recent years, a rich body of research has been conducted to explore the latent psychological mechanisms that systematically explain how technological attributes impact consumer behaviors. Existing research can generally be organized into the S-O-R behavioral framework \cite{ahmed2022social}, as shown in Fig.~\ref{Visio-fig1} (Part I). Technological attributes, i.e., the stimulus (S), triggers certain cognitive or affective shifts within consumers (O), which consequently leads to behavioral or attitudinal responses (R). For example, if the MSP invests funds to improve the comfort of VR hardware and increase the types of services in the virtual world (System Design), users can directly feel a smooth interactive experience and enjoy informative services (S). The users will then consider that these Metaverse services are useful (O), resulting in brand loyalty to the MSP and willingness to promote the services (R).

Specifically, in the context of VR/AR, as the core access to Metaverse, {\textit{stimulus}} constructs can be grouped into three clusters. The first cluster includes the media attributes, such as the perceived interactivity, the perceived vividness, and the perceived personalization \cite{kumar2021augmented,du2022attention} to the virtual tourism attraction. The second cluster covers the media quality, including the richness of information and system response time \cite{kumar2021augmented}. Finally, the third cluster in the stimulus concerns the ergonomic design of equipment \cite{tsao2018human}. Despite being the antecedent of consumer behaviors, the stimulus serves as direct outcomes of system designs. For example, the resource allocation decision can directly impact the media quality perceived by consumers. Such a dual property explicitly bridges Metaverse system design considerations with consumer behaviors, which well supports applications of the interdisciplinary framework.

Receiving the stimulus, consumers start to form psychological evaluations of the system technologies, which is called {\textit{organism}}. The evaluation can be specified as cognitive attitudes such as expected utilitarian value from technologies and affective attitudes including enjoyment \cite{kumar2021augmented}. Meanwhile, the authors in \cite{feng2019privacy} suggest that privacy concerns and worries about loss of control over information access may arise, which can hinder the adoption of cutting-edge technologies. 
Finally, the psychological organism results in behavioral {\textit{responses}} including purchase, Word-of-Mouth recommendations to attractions, and attitudinal responses such as brand loyalty to MSPs \cite{kumar2021augmented}.

\subsection{Connect System Design with Consumer Behaviors}\label{S32}
Although the S-O-R causal chains can connect the Metaverse system designs with consumer behaviors, the establishment of a quantitative relationship is a non-trivial task. Researchers have to solve explicitly the following two questions:
\begin{itemize}
\item[{\textbf{Q1)}}] How can the psychological states described in Section~\ref{S31} be reliably measured?
\item[{\textbf{Q2)}}] How can we connect consumers' psychological states with attributes that can be directly applied to Metaverse system design processes?
\end{itemize}
\subsubsection{Structural Equation Modelling}
To address {\textbf{(Q1)}}, the Structural Equation Modelling (SEM) can be used, which covers a family of causal inference methods~\cite{pal2018analyzing}. The SEM allows systematical and reliable measurements of the latent constructs, i.e., the psychological states when users participate in the Metaverse services. By considering that the latent constructs are the causes of observable measurements, e.g. users' answers to a Liker-scale survey questions designed by MSP, the latent constructs can be inferred. To ensure the reliability of the inference, several observable measurements should be simultaneously applied for one latent construct. In addition, the level of variance commonality across the observable measurements can be examined to validate the assumption that the selected observable measurements are ``caused'' by the same latent construct.

\begin{figure*}[t]
	\centering
	\includegraphics[width = 1\textwidth]{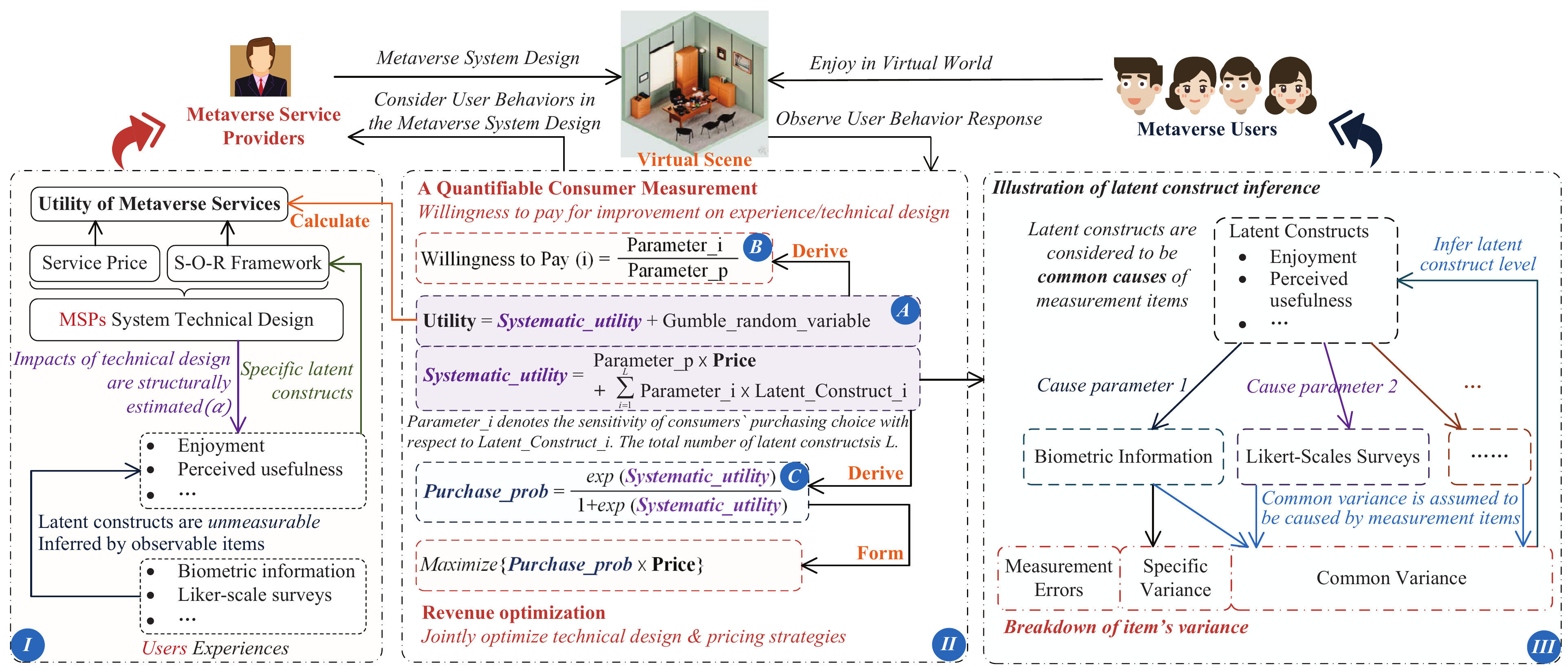}
	\caption{Detailed illustration of Metaverse system-consumer integration.}
	\label{Visio-fig4} 
\end{figure*}
The detailed progress discussed above can be illustrated in Fig.~\ref{Visio-fig4} (Part III). For example, to infer the latent constructs such as user enjoyment in the Metaverse, researchers can collect several observable measurements including biometric data, such as eye concentration. The variance of a measurement can be decomposed into several factors, e.g., measurement error, variance specific to this measurement, and the common variance caused by latent constructs. With the common variance, the causal parameters between observable measurements and latent constructs can be derived. Consequently, latent constructs can be reliably inferred.


\subsubsection{Econometric Discrete Choice Modelling}
Although SEM can provide quantitative evidence for latent causal relationships, such scale-based numerical parameters can hardly be interpreted meaningfully for Metaverse system design. Hence, it is vital to explicitly connect latent constructs with practical attributes, for instance, the expected revenue. As illustrated in Fig.~\ref{Visio-fig4} (Part II), the DCM can provide a paradigm to link psychological constructs with observable attributes {\textbf{(Q2)}}. The DCM is based on the Random Utility Model (RUM), which considers the Metaverse users choose the option that maximizes their utility. The utility is composed of a systematic portion, which is calculated linearly with identified attributes, and a random portion \cite{train2009discrete}. Specifically, different latent constructs, e.g., service enjoyment and usefulness, have different weights on different users. We call the weight as the parameter to the corresponding latent construct. Then, the systematic utility can be obtained by adding latent constructs weighted by corresponding parameters, taking into account the price and its weight. The utility function can finally be expressed as the addition of the systematic utility and a random variable.
Furthermore, through integration over the probability density function of the random portion, the probability of consumers choosing a certain option can be analytically expressed as functions of systematic utility in binomial/multinomial formulations, as shown in Fig.~\ref{Visio-fig4} (Part II-C). On the user population level, such probability can also be interpreted as the market potential for a product. With observations of consumers' actual choices, e.g., to use Metaverse service or not, the structural parameters of systematic utility can be estimated through methods such as maximum likelihood and methods of the moment. Managerial implications can be readily obtained with the estimated structural parameters.

\subsubsection{Insights}
First, with the estimated structural parameters, researchers can readily derive the sensitivities of Metaverse consumers' purchase probability/market potential with respect to various attributes. For example, to understand consumers' responses to price adjustment, the MSPs can take the derivative of purchase probability with respect to price. Here the price could be the Metaverse service subscription fee. The resulting numerical values can provide insights into the potential market shrinking because of the increase of unit price.  

Second, the DCM framework acknowledges the MSPs about consumers' trade-offs between different pairs of attributes. One important economic indicator is the ratio between the price parameter and any other parameter, a.k.a, the willingness to pay (WTP), as shown in Fig.~\ref{Visio-fig4} (Part II-B). In this way, DCM allows the monetization of psychological shifts and directly connects improvement in consumers' satisfaction with system designs through the WTP measurement \cite{vij2016and}. For example, the MSP can quantitatively determine whether to invest more resources to improve a system design indicator, by calculating the ratio of the price increase due to more resource input and the latent construct increase brought by the improvement of system designs. If the ratio, i.e., WTP, is calculated to be greater than 1, this decision of the MSP is reasonable.

Compared to the existing QoS or QoE measurements, the DCM framework provides a macro-level path for implementing consumer-centric approaches. The MSPs can explicitly model the expected revenue as the product of Metaverse service purchase probability and price, e.g., service subscription fee. Then, the revenue model can be readily embedded into the objective function for the Metaverse system design. In addition, the consumers' purchase probability is modeled as a function of Metaverse service attributes, which are directly determined by service costs. For instance, the probability of users purchasing Metaverse services can be formulated as a function of VR media quality, which is directly associated with the MSPs' available bandwidth and modeling ability.
Compared to the existing system design method which optimizes the technical indicators only, the integrated framework leads to the overall system optimality, bringing more profits to MSP.

\subsection{Heterogeneity in Metaverse Users}\label{S33}
In real-world Metaverse applications, it is significant to consider the interpersonal and contextual differences among users, as known as heterogeneity. In the case of Metaverse, for instance, attitudes toward privacy can vary significantly across different socio-demographic groups. For example, the authors in \cite{lee2019information} conclude that individuals who are male, younger, earn higher incomes, and hold higher education qualifications are more sensitive to their information privacy. Hence, from the WTP perspective, it implies such consumer group needs to be compensated more for the collection of personal data. For example, if the MSP plans to analyze the user's attention by collecting the eye movements of users to design a personalized resource allocation scheme~\cite{du2022attention}, it is necessary to provide more compensation for such privacy-sensitive users or consider alternative schemes.
Note that the heterogeneity is not only attributed to interpersonal differences. The usage contexts may also differentiate consumers' behaviors. For example, in a working context such as Metaverse conference, informativeness is the dominant factor for technology adoption. While in a virtual traveling context, the authenticity of Metaverse users experience prevails~\cite{kim2020exploring}. 

Hence, at the design stage, a homogeneous consumer measurement may not suffice to deal with real-life Metaverse system design complexities. One straightforward solution is the incorporation of latent constructs into the DCM that is discussed in Section~\ref{S32}. As a result, the interpersonal heterogeneity among Metaverse users arising from attitudinal differences can be captured \cite{li2019product}. In Section~\ref{S32}, we treat the structural parameters in the DCM to be deterministic. To capture the interpersonal heterogeneity, econometrician has developed more advanced DCMs by treating the structural parameters as random variables. Such models include the Mixed Multinomial Logit model (MMNL), which assumes structural parameters to be continuously distributed, and the Latent Class Model (LCM), which assumes a discrete distribution \cite{train2009discrete}.
These methods have great potential to be applied in the Metaverse system design as an alternative to the DCM in our proposed framework.

\section{A Metaverse Service Design Case}\label{S4}
In this section, we consider a Metaverse service design case, and show how the proposed integrated framework can facilitate consumer-centric decision makings.
\subsection{Design Case Introduction}\label{design}
Here we consider the virtual traveling as the Metaverse service case. Specifically, users access the Metaverse with devices such as VR and AR. Then, users are free to choose the scenes of the virtual traveling, such as a digital copy of an attraction or a completely virtual place. The MSP needs to ensure multiple system technical indicators such as~\cite{van2022edge}
\begin{itemize}
	\item Accurate Sensing Ability: To create a digital copy of the real world, the MSP needs to sense the real world through various sensors. The inauthenticity caused by inaccurate sensing leads to a degraded service experience.
	\item High Downlink Data Transmission Rate: To send a large amount of virtual object data to users, the MSP needs to ensure a high downlink data rate. Untimely data transmission causes slow loading of Metaverse scenes.
	\item Low Latency: To provide a realistic travel experience in real time, the MSP needs to ensure low latency. High latency creates jitter in visual effects.
	\item Hardware Design: To avoid problems such as heat and excess weight of wearable devices, the MSP needs to invest in hardware design. Uncomfortable hardware design leads to adverse reactions such as motion sickness.
\end{itemize}

\subsection{Psychological Causal Chain for Virtual Travel}
To adopt the S-O-R causal framework described in Section~\ref{S31}, the MSP should first formulate the mechanism behind system design and consumers' reactions. In the virtual traveling setting, perceived authenticity of the travel experience is regarded as a significant stimulus (S). It is defined as the perceptions of the extent to which the experiences are real, original and exceptional. Then, the perceived authenticity leads to consumers' enjoyment and emotional involvement in the Metaverse as the psychological organism (O). Consequently, consumers' purchase intention can be induced \cite{kim2020exploring} as the response (R) to the organism. As discussed in Section~\ref{S32}, the SEM allows the estimation of causal parameters along the S-O-R framework. For example, with a scale increase in the perceived authenticity, the MSP can estimate the magnitude of increase in consumers' enjoyment, emotional involvement in the Metaverse, and the purchase intention to the virtual goods that may be involved in virtual traveling. Furthermore, to establish connection between system design with the psychological process, the dual property of stimuli constructs shall be addressed as described in Section~\ref{S31}. The MSP can adopt a regression method to estimate the impacts of technical indicators on the level of perceived authenticity in the virtual traveling. Consequently, a complete technical-to-response causal chain can be established.

\subsection{Consumer-Centric System Design Optimization}
As noted in Section~\ref{S32}, though the SEM method enhances MSP's understanding on Metaverse consumers' behaviors, results are hardly to be applied directly to system design process such as virtual attraction creation. The reason is that all parameters estimated under the SEM method are ``scale'' based with no direct practical explanation. Thus, we explicitly describe a consumer-centric service design optimization progress with the application of the DCM part in our proposed integrated framework. We first start with a formulation with technical indicators being directly applied in the DCM equation. Although such an approach omits the psychological mechanism behind consumers' behaviors, it allows the joint optimization in a more straightforward and easy-to-implement manner. Following that, we incorporate the latent constructs, e.g., the enjoyment users obtain from the virtual traveling and the perceived usefulness of the service, into the Metaverse system design progress.

\subsubsection{Direct Applications of DCM with Technical Indicators}
As illustrated in Section~\ref{S32}, the measurement of latent constructs can be a non-trivial task, which requires to collect additional panel surveys or biometric data. In cases when such data is unavailable, MSPs can directly incorporate the relevant technical indicators as we discussed in Section~\ref{design} into the DCM-based utility formulation as shown in Fig.~\ref{Visio-fig4}. The systematic utility can be expressed as a linear function with respect to all technical indicators, with a set of structural parameters to quantify the impacts of each indicator on Metaverse consumers' choices. Specifically, these structural parameters can be estimated by recording consumers' choices of virtual traveling adoption under different technical designs.

\begin{figure}[t]
	\centering
	\includegraphics[width = 0.48\textwidth]{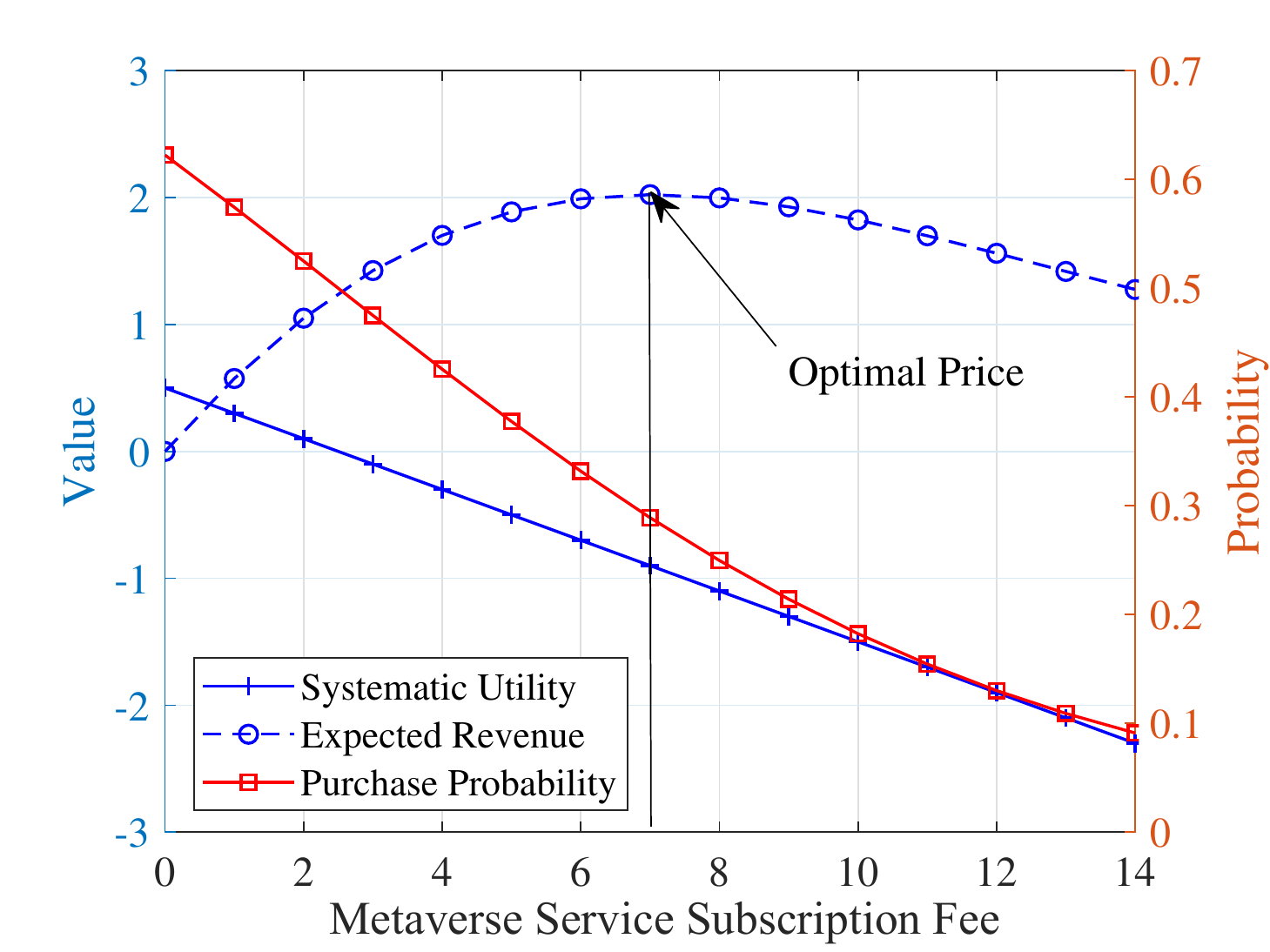}
	\caption{An illustration of shifts in consumer utility, purchase probability and revenue under different Metaverse service subscription fee.}
	\label{Visio-price} 
\end{figure}

With the estimated structural parameters, a joint optimization problem can be formulated to maximize the product of the purchase probability and the Metaverse service subscription fee, i.e., the price, as shown in Fig.~\ref{Visio-fig4} (Part II-C). Consequently, the optimal set of product attributes, including technical indicators and pricing, can be determined. An example is shown in Fig.~\ref{Visio-price}, after obtaining the parameters of the latent constructs of Metaverse users (Here we show the results of the simulation), the MSPs can analyze the shifts in consumer utility, purchase probability, and revenue under different service subscription fee. The best pricing strategy can then be determined.

Furthermore, our proposed framework can also support MSP's service differentiation scheme. For example, the MSP intends to establish a premium virtual traveling service with higher data transmission rate. By substituting the new services' technical details into the DCM, the MSP can estimate the potential market size of the new service offers. Meanwhile, through investigating the ratio of the structural parameter of data transmission rate to that of service price, the MSP can derive consumers' willingness to pay for incremental data transmission rates for a smoother virtual traveling experience. Such WTP measurement can well-support the pricing scheme for the premium services.

\subsubsection{A Full Application with Latent Constructs}
Although the direct incorporation of technical indicators can suffice to provide meaningful practical implementations, the incorporation of latent constructs can bring several additional benefits. First, from the statistical perspective, it has been shown that such incorporation can improve the forecast accuracy of the DCM. Also, the variances of estimations can be reduced, and the biases can sometimes be eliminated \cite{vij2016and}. Second, by opening up the ``black-box'' relationship between technical indicators and Metaverse consumer behaviors, the explainability and transparency of consumer modeling can be improved. For example, when only technical indicators are incorporated, the MSP may find that the low bit error probability leads to greater purchase probability yet limited consumer-related insights can be drawn. In contrast, if the latent constructs are adopted, the aforementioned relationship can be explained with the ``{\textit{bit error probability}} $\to$ {\textit{authenticity in virtual traveling service}} $\to$ {\textit{purchase probability}}'' causal chain. In a scenario when the authenticity is not a priority, the MSP can readily de-prioritize the bit error probability indicator as well.

\section{Future Research Directions}\label{S5}
\subsection{Advertising Strategy of Metaverse Companies}
The advertising can be designed to attract new Metaverse users and encourage existing users to make multiple purchases, for increasing the sale of Metaverse services or digital products. Moreover, advertising strategies can be used in the allocation scheme of network resources, i.e., bandwidth, required by users to access the Metaverse. Specifically, users could be incentivized to buy more network resources to increase the Metaverse access rate. Considering the total budget of a MSP is limited, the allocation of the budget between advertising and network resources needs to be optimized. The competition between multiple Metaverse companies is also worth studying.


\subsection{Social Network Analysis}
As noted in Section~\ref{S4}, the successful implementation of the consumer-centric design depends on the knowledge to consumers, which requires experiments among consumers. However, the experiment design is complex. For example, to estimate DCM structural parameters, the MSP may invite users to experience pre-defined Metaverse environments and record their choices. The environments should be defined in a way that much information can be revealed. If in one experiment, all technical indicators are set to the highest standard, consumers are likely to adopt the service and little information can be obtained from such a experiment design. To facilitate this process, interdisciplinary collaborations with social scientists may be pursued.

\subsection{Dataset Study for Metaverse}
Because the research of Metaverse is still at an early stage, there is not enough datasets to meet the needs of various system design schemes. Therefore, an important research direction of Metaverse is the establishment and analysis of datasets. For example, datasets related to virtual scenes can help us to analyze users' behaviors to design real-time resource allocation schemes to improve QoE, and datasets of consumer responses to various strategies of MSPs can help MSPs formulate suitable market solutions.
\vspace{1cm}

\section{Conclusion}
We have proposed an integrated research framework that encompasses both consumer behaviors and system design considerations for the MSPs, which facilitates the consumer-centric Metaverse ecology development. Under the proposed framework, the S-O-R causal chain was introduced to bridge systematically technical indicators with consumers' responses. By rethinking QoE measurement and analyzing the underlying causal chain, a micro-level WTP measurement and a macro-level DCM-based optimization formulation were introduced. Compared to the current Metaverse system design solutions that only focus on QoE, our proposed framework systematically describes the relationships between technical design, consumers' utility and profitability for the MSPs. It allows a joint optimization with more transparent and explainable process. Meanwhile, a virtual traveling design case is introduced to operationalize the proposed framework. This article also aims to advocate the wider adoption of interdisciplinary perspectives for the study of increasingly diversified and decentralized next generation Internet. Efforts shall be made to synergies knowledge from the social science track into the futurist system design.

\bibliographystyle{IEEEtran}
\bibliography{ref}

%
%
%
%
 
\end{document}